# An Energy Driven Architecture for Wireless Sensor Networks

Doan B. Hoang
INEXT Centre for Innovation in IT Services and Applications
University of Technology, Sydney
+61 2 95247943

dhoang@it.uts.edu.au

Najmeh Kamyabpour
INEXT Centre for Innovation in IT Services and Applications
University of Technology, Sydney
+61 2 95244609

najmeh@it.uts.edu.au

*Abstract*-**Most wireless sensor networks operate with very limited energy sources-their batteries, and hence their usefulness in real life applications is severely constrained. The challenging issues are how to optimize the use of their energy or to harvest their own energy in order to lengthen their lives for wider classes of application. Tackling these important issues requires a robust architecture that takes into account the energy consumption level of functional constituents and their interdependency. Without such architecture, it would be difficult to formulate and optimize the overall energy consumption of a wireless sensor network. Unlike most current researches that focus on a single energy constituent of WSNs independent from and regardless of other constituents, this paper presents an Energy Driven Architecture (EDA) as a new architecture and indicates a novel approach for minimising the total energy consumption of a WSN.**

## I. INTRODUCTION

Sensors have been used to some extend in various industrial applications in past decades, however, their use only took off massively when they acquire the wireless capability, able to form networks with neighboring sensors, and are miniaturized. With the advance of electronics and the miniaturization technologies, wireless sensors and wireless sensors networks (WSNs) have found applications in many fields including environment, health, disaster alert, car, building, and mining industries and they have the potential of revolutionize many aspects of our lives.

It is also recognized that sensors are most useful when they are deployed in large numbers, especially for collecting environmental map of a geographical area such as a complete building, an agriculture field, or a rain forest. They are crucial in many critical applications where they are deployed in inaccessible and dangerous areas to collect data related to the environment, battlefields, or nuclear reactors, etc. Once the sensors are deployed, they may no longer be accessible for further physical manipulation such as fixing faulty components or changing batteries. Furthermore, their operation (power emission) must not interfere with the environment (e.g., aircraft operation) or cause harmfully to people.

The fundamental question is how to design wireless sensor networks with extended lifetime long enough to provide useful information efficiently, and cost effectively. Realizing that sensors must consume energy and must work collaboratively to deliver data as dictated by the application, the challenge is how to minimize the energy consumption of the whole sensor network taking into account of various constraints of the application.

To provide answers to this question and its constraints one needs to address many challenging problems:

*Scalability*: The whole system should not be overloaded with raw data as sensors can sense and collect data automatically and frequently and hence potentially generate extremely large volumes of data over time and consume excessive amount of energy.

*Reliability*: Sensors must produce reliable data as they are critical data in many applications. It is a challenge to design reliable protocols at the same time minimize the energy consumption overheads.

*Collaboration*: Sensors can only send data a short distance away from itself. They must collaborate effectively with their neighbors to carry out the functions required by the application.

*Security*: This is an essential aspect of a WSN and it should be an integral part of any design from the outset.

Clearly, to minimize the energy consumption of wireless sensor networks, many inter-related factors must be considered. For example, the pattern of energy consumption (and hence the quantity) of an individual sensor is often dictated by the goal of the application. To deliver its data, a sensor has to rely on its neighbors to relay its data to the destination and the way sensors form their interconnected networks certainly play a crucial role in determining the energy consumption of the overall networks and application. Sensing mechanisms, transmission mechanisms, networking protocols, topology, and routing all play crucial part in the overall energy consumption and they are often interrelated as part of a complex system and this makes it difficult to analyze or optimize.

Efforts in minimizing energy consumption have increased over the last few years, however, they mostly focused on some specific and separate components of energy dissipation in WSNs such as MAC protocols [1], [2], routing [3], topology management [4] and data aggregation [5]. These components are, however, highly integrated within a WSN but they their interplay cannot be taken into account as each constituent is treated independently without regard for other constituents. Minimizing the energy consumption of one constituent may increase the energy requirements of other constituents and hence may not guarantee the minimization of the overall energy consumption of the entire network.

There is a real need for a unified framework where all major constituents of a WSN are brought under one roof so that the interplay among the constituents can be taken into account in optimizing the level of energy consumption of the whole network.

In this paper, Energy Driven Architecture (EDA) is proposed as a general and novel approach. One significant feature of EDA is the introduction of WSNs as constituent-based energy systems. The result is a constituent-based network architecture, that enables new approach in energy optimization of WSN and that allows existing approaches to be adapted to this architecture.

Network architectures such as OSI and Internet are basically functional models organized as layers with the layer below provides services to the layer above and eventually the application layer provides survives to the end users. Network is often evaluated in terms of its quality of service parameters such as delay, throughput, jitter, availability, reliability and even security. However, when it comes to energy consumption, one often encounters difficulty in evaluation and hence optimization as there hardly exist any models that take energy consumption into account. As discussed in the related work of section 4, researchers fall back to the traditional network architecture and try to minimize selected component of a single layer with the hope that the overall energy consumption of the network is reduced without regard for other components or layers. This is hardly an ideal situation where one does not know how a single component fits within the overall energy picture of an entire wireless sensor network.

The rest of the paper is structured as follows. We first introduce the new energy driven architecture and its constituents in section 2, followed by a discussion on the overall architecture in section 3. Some related efforts in minimizing individual components of WSNs are summarized in section 4. Finally, we summarize our work and outline future research directions in section 5.

## II. PROPOSED ARCHITECTURE

We propose in this paper a totally new model where energy is the focus. The model is called EDA. In this model, we identify major energy consumption components in terms of their roles/activities relative to the network and with respect to the application and model the whole wireless sensor network accordingly.

Starting from an individual sensor, without concerns for its neighbors, without concerns for the network, it has to spend part of its energy to be alive and function (e.g., capturing sensed data). But a sensor does not live alone, it has to interact with its neighbors and interact with the local community, hence it has to spend part of its energy to maintain its neighborhood interaction. Further afield, a sensor cannot fulfill its duty without establishing a communication channel to transport its data to the destined sink as required by the application. To do so it has to establish network(s) and other sensors. Clearly, all sensors have to collaborate in establishing and maintaining network(s) of mutual interests. The topology of the network, the routing protocol, and other aspects are often dictated by the objective of the application with various constrains in terms of performance and the environment.

Taking the above discussion into consider, Energy Driven Architecture (EDA) is proposed based on five general energy consumption constituents of sensors in WSNs (figure 1): Individual, Local, Global, Sink, and Environment. The individual energy consumption involves all basic sensor operations that allow it to exist: processing, storage and querying in/from memory, sensing and digitalize signals and convert a sequence of bytes to and from radio waves.

The local communication is concerned with initiating and maintaining all communications between a sensor node and its immediate neighbors so that they can co-exist to perform the roles within the WSN as dictated by the objective of the application. There are several reasons that nodes consume energy at this level. Sensors must establish at least a pathway through neighbours to forward their data to the destination. Sensors have to be ready for responding to requests.

Neighbor monitoring is a necessary and costly function. Nodes should be aware of the current available resources of their neighbor's such as residual energy and channel state information [6, 7], memory space, etc. Based on this information, sensors are able to make a good energy-saving decision in choosing an appropriate neighbor-node for relaying their packets. In case where networks support mobility, it is necessary that nodes update their neighbor information in an efficient manner.

The global communication is concerned with global strategies for maintaining the whole sensor network and for transporting all sensors' data to the destination, the sink. Selecting relevant network topologies, choosing efficient routing methods become a major consideration. Adopting a network topology may depend on the objective of the application. Routing methods help minimize the number of relay hops to the destination which are expensive in terms of energy dissipation. Furthermore, inappropriate topology and routing may create congestion and packet loss hence increasing energy consumption of the network.

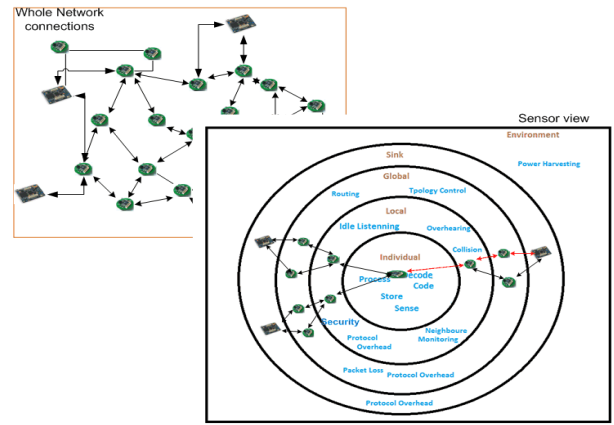

Figure 1. Energy Driven Architecture

Sensors may be able to harvest energy from their environment and this is a positive effect on their total energy. Increasingly, this feature becomes extremely important as it has the potential to sustain the WSNs until the end of their useful lifetime. However, it may impose additional complexity and costly operations on topology management and routing protocols.

The sink is a powerful component and can operate like a manager in the network. It plays an important role in balancing management, control, data collection and energy minimization of the whole sensor network. For example, it may reduce number of control packets by sending control information such as initial topology and routing information via beacon nodes. This eliminates energy-costly operations that are supposed to be performed by all nodes.

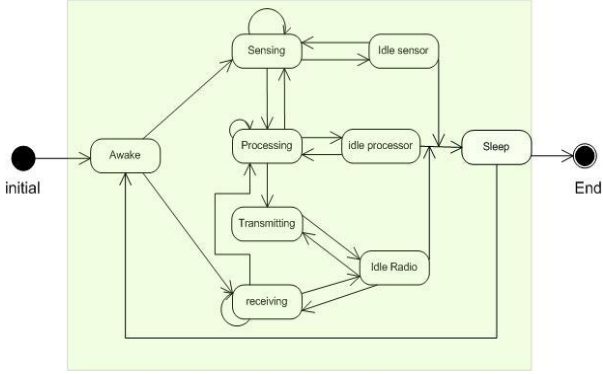

Figure2. General State Diagram for sensor nodes.

HEPA stipulates an energy constituent-based and general approach and can be used to deploy WSNs based on a minimizing overall energy consumption viewpoint.

We consider energy consumption of each constituent in a time interval $\Delta t$, $E(\Delta t)$ as follows:

$$E(\Delta t) = \frac{\partial E}{\partial t}\Delta t \qquad (1)$$
$$\Delta t = t_2 - t_1$$

In the following sections each constituent is discussed in more detail.

*A. Individual Constituent*

The "Individual constituent" consists of five main controllable and programmable units: the sensing unit, the processing unit, the memory unit, the radio unit, and the power supply unit. Together, these units perform all the essential and basic operations for the sensor to just exist. The processing unit executes instructions and processes data. The memory unit deals with storage for data and instructions. The sensing unit gathers analog signals from environment and converts them to digital signals for processing unit. The radio channel digitalizes radio waves and converts a stream of bytes to radio waves. Since these units consume different amount of energy in active, sleep and idle states, a state-based scenario is often assumed for modeling energy consumption in Individual constituent (Figure 2).

According to figure 2, switching from the initial state to the awake state consumes energy as it involves loading and executing instructions. After an initialization, a node is prepared to generate or receive data. The received and generated data are processed by the processor to decide when and where the data should be sent or how sensors behave. All these operations are done in an active state and all units move to the idle state when they do not have to perform any task. Even in the idle state, sensors still waste some amount of energy because of leakage current. To preserve energy, relevant circuitry should be switched off.

Moreover, switching among the unit's states also consume considerable amount of energy, so number of switchings should be reduced. To reduce the level of energy consumption optimally, some minimizing algorithms should dictate the ways operations of different units are performed. For example, the processor should minimize the amount of storage and memory queries, and memory read and writes operations should also be done in an energy efficient manner. The energy consumption of the individual constituent of node $i$ in a time interval $\Delta t$ can be formulated as:

$$E_{individual_i}(\Delta t) = \sum_{u=1}^{N_u}\sum_{s\in S}\sum_{w\in W} I(e_{u,s}, e_{u,w}, t_{u,s}) \qquad (2)$$

Where $N_u$ is number of units and $U$ is:

$$U = \{Pu, Su, Mu, TRu\} \qquad (3)$$

$U$ is a set of individual units where $Pu$ represents the processing unit, $Su$ the sensing unit, $Mu$ the memory unit, and $TRu$ the transceiver unit for digital signal processing. $S$ defines a set of sensor states as follows:

$$S = \{sleep, awake, active, idle\} \qquad (4)$$

and a switching transition, $w$, is defined as follows:

$$W = \{is, sw, wa, ai, ia\} \qquad (5)$$

Where $w \in W$ and $W$ is a set of possible switching transitions of each component, *is*: idle to sleep, *sw*: sleep to awake, *wa*: awake to active, *ai*: active to idle, and *ia*: idle to active. Therefore $e_{u,s}$ shows the energy consumption of unit $u$ in state $s$ and $e_{u,w}$ shows the energy consumption of unit $u$ for a switching transition defined in $W$. $t_{u,s}$ define duration of states for unit $u$.

*B. Local Constituent*

Generally a local constituent deals with initiating and maintaining all communications between a node's immediate neighbours. The local constituent consumes energy in following ways to perform application-dependent roles:

- Neighbour monitoring for gathering information of neighbour's available resources such as residual energy and memory space. The gathered information can be used for topology management, routing and mobility management [8].

- Security management for preventing malicious nodes from destroying the connectivity of the network and tampering with the data. Malicious nodes can manipulate and drop the exchanging packets through the network. In local level each node, may use security protocols for distinguishing malicious nodes in their covered area and remove the connection with them.

Pertaining to the local constituent, other activities that consume energy may include:

- Idle listening - if the node's antenna does not receive or send a message, it remains on listening mode while nothing happens, but it still consumes some amount of energy.

- Collisions management - if the node does not receive acknowledgment of the transmitted packet, it has to retransmit the packet. This situation happens when neighbours transmit packets on the shared medium at the same time.

- Overhearing - the node receives packets that are sent to the shared medium and they are not destined for it, however, it still has to examine the packet to figure out what to do.
- Local communication protocols - various local communication protocols have to be performed to maintain the node's relationship with its neighbours. This type of protocol overheads must be taken into account in terms of energy consumption.

The following equation summarises the local energy consumption of node $i$ in a time interval $\Delta t$:

$$E_{local,i}(\Delta t) = \sum_{j \in neighbour} L(e_{ij}(mon), e_{ij}(sec), e_i(idle),$$
$$e_{ij}(local), e_{ij}(coll), e_i(ohear)) \quad (6)$$

Where $j$ is a member of node $i$ neighbours and the local energy consumption, $E_{local,i}$, in a time interval $\Delta t$ can be expressed as a function of several energy consumption components[9]:

- Energy consumption for neighbour monitoring, $e_{ij}(mon)$. This component is determined based on the number and the size of exchanging packets required for determining available resources of the neighbours in ideal situation.
- Security protocol, $e_{ij}(sec)$. This component is determined based on the size and the number of packets for authenticating and authorizing neighbours.
- Energy consumption of idle listening, $e_i(idle)$. This component is determined based on energy consumption of the leakage and the duration of idle listening.
- Local protocol overhead, $e_{ij}(local)$. This component is determined based on size and number of control packets.
- Collision, $e_{ij}(coll)$. This component is determined based on the size and the number of retransmitted packets.
- Overhearing, $e_i(ohear)$. This component is determined based on the size and the number of listened packets with different destinations.

*C. Global Constituent*

The global constituent is concerned with the maintenance of the whole network, the selection of a suitable topology and an energy efficient routing strategy based on the application's objective. This may include energy wastage from packet retransmissions due to congestion and packet errors. The global constituent is defined as a function of energy consumption for topology management, packet routing, packet loss, and protocol overheads. The energy consumption of the global constituent in a time interval $\Delta t$ can be formulated as:

$$E_{global,i}(\Delta t) = G(e_i(topo), e_i(route),$$
$$e_i(global), e_i(pktls)) \quad (7)$$

Where $e_i(topo)$ represents the energy consumption for establishing a relevant topology through the nodes based on the application's objective.

$e_i(route)$ represents the energy consumption for determining and maintaining hops and transporting packets to the destination. The number of relaying hops can be expressed as a cost component in term of energy dissipation. It should be determined and minimized by a suitable routing method. The cost for maintaining the network connectivity should also be accounted for if hops fail during the network life time.

$e_i(global)$ represents the energy consumption due to protocol overheads. It is calculated based on the cost transporting control packets for maintaining the overall network topology and configuration.

$e_i(pktls)$ represents the energy consumption due to packet loss. Selecting inappropriate topology and routing methods may cause congestion and packet-loss in the network. In this case, extra energy consumption has to be added if a node is required to retransmit a packet.

*D. Environment Constituent*

In cases where nodes are capable of extracting or harvesting energy from the environment, we propose to take into account this positive energy component in determining the lifetime of the WSN. However, deploying this harvesting energy capability may also incur extra energy consumption in other constituents as the activity entails switching between states. For example, if residual energy can not be anticipated because of using energy harvesting nodes, it may cause extra energy consumption in the group constituents. The environment constituent as a positive energy component can be formulated as follows.

$$E_{battery,i}(\Delta t) = -H_i(t) \quad (8)$$

Where $H_i(t)$ is amount of harvested energy in a time interval $\Delta t$.

*E. Sink Constituent*

The sink(s) often assumes the roles of manager, controller or leaders in WSNs. The sink constituent represents the component that consumes energy to direct, balance and minimize the energy consumption of the whole network and to collect the generated data by the network's nodes. Using the sink in that manner can eliminate energy-costly operations of all nodes. For example, in the network establishment stage, the sink sends control information based on the application's objective to the whole network to ensure that all nodes execute some cost-saving measures. Energy consumption of node i in a time interval $\Delta t$ from the sink constituent viewpoint can be formulated as follows:

$$E_{snk,i}(\Delta t) = K(e_i(snk)) \quad (9)$$

Where $e_i(snk)$ shows consumed energy of node $i$ to communicate with the sink and perform sink's commands.

## III. EXPERIMENTAL RESULT

In this section we used Energy Driven Architecture to compare two approaches. Also we show how we can increase the performance of an approach by changing the effective parameters of EDA's constituents.

Since in reality the energy consumption of constituents is not observable we consider lifetime of the sensor as a measure to compare two routing approaches. In this experiment the lifetime of a typical sensor is monitored.

Two routing approaches are assumed to transport data from sensors to the sinks. First approach is called Selective; nodes

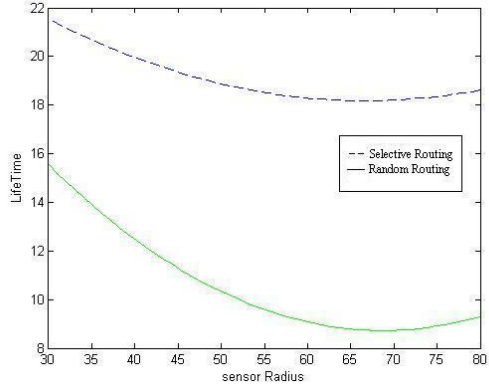

(a)

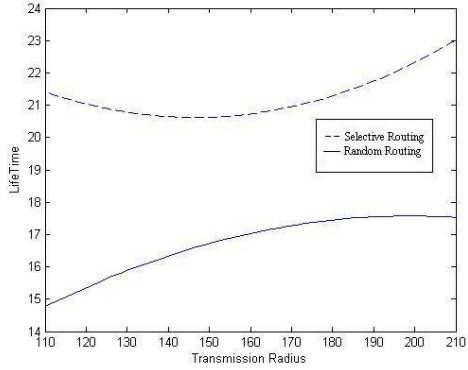

(b)

Figure 3.Lifetime of a typical sensor with different routing method in a Wireless sensor network (a) optimum lifetime by change an effective Individual parameter (sensing radius) (b) Optimum life by change an effective Local parameter (Transmission Radius)

select a neighbor to relay data based on neighbor's residual energy and busy degree. Second approach called Random, a neighbor is selected randomly.

In the following, we compared these methods in term of packet overhead of different constituents:

$$\begin{cases} b_{Global,R} > b_{Global,S} \\ b_{Local,R} < b_{Local,S} \\ b_{Individual,R} < b_{Individual,S} \end{cases} \quad (10)$$

Global constituent's tasks have the highest cost for the sensor in compare with Individual and local packets. In this experiment Selective method increase Individual tasks (e.g. process of incoming packets) and local tasks (e.g. neighbor monitoring to collect information about the neighbor's residual energy). On the other hand the random method loads a massive number of global packets to the node without considering the node status. The node may select more often and the rate of packet loss may go up and this loads extra global communication on the sensor. As it can be seen in figure 3 the sensor with Selective method generally has longer lifetime. As a result, it is worthwhile if we decrease global overhead and load works on local and individual constituents.

In order to increase the performance of these methods we assumed different values for effective parameters of individual and local constituents. Figure 3 shows the optimum value of two effective parameters of constituents which results longer lifetime by using different routing approaches.

## IV. DISCUSSION ON THE OVERALL EDA

Our overall goal is to derive a formulation for the total energy consumption of the whole wireless senor network in order to optimize and design energy-efficient WSNs. This drives us towards an energy-functional approach where by a node within a WSN sees itself occupying a number of roles, each of them requires the consumption of a portion of the overall energy of the whole network: an individual, a member of a local community, a member of a global community (the whole network), and even a manager of the intended application. In doing so, we arrive at the novel energy-constituent based EDA architecture.

Modeling EDA constituents as single energy consumption units within a hierarchical structure presents many possible strategies for maximizing the network's lifetime. Viewing a WSN as a composition of energy-consuming constituents has a number of benefits. It clearly shows how energy of a node is consumed by tasks, operations, events, changes, demands and commands during its lifetime. It allows the optimization of the energy consumption of a node if desired. It allows the optimization of a selected constituent for a specific application. Most importantly, it allows an overall optimization of the energy consumption of the entire network by considering the play-off among constituents. Furthermore, the architecture is robust and flexible in that each constituent can be adapted to suit the required application.

EDA offers a flexible energy-based model for all types of sensor applications. The individual constituent represents the controllable and programmable part of sensors that consumes energy for executing, generating, and interpreting code and data which is loaded by the local, global, environment and sink constituents into a sensor or even if the node is idle. The local constituent represents components that consume energy for tasks and events around node to allow it to exist among its neighbors. The global constituent represents the part that consumes energy essential for the existence of the whole network as it deals with the organization and communication among all nodes in the network. The environment constituent presents an energy harvesting opportunity. The sink constitutes a control component that helps distribute and direct the overall network operation to achieve the goals of the intended application.

It is clear that EDA only presents an architecture that expresses the essential energy constituents and their relationship in a wireless sensor network. The task of formulation of a single integrated overall energy consumption of the system (WSN and its applications) remains to be explored. Several approaches for such formulations are described in the conclusion and future work.

It should be noted that a WSN is composed of constituents and interactions among constituents that are essential for the operation of the whole network and its application. Completely separable constituents are not always realistic, some degrees of overlapping are essential. For example, the global constituent and the local constituent seems overlapped as they use a common energy consuming component, the radio channel, but they consume energy for different roles and purposes. For this reason, optimizing individual constituents only produce suboptimal results. A complete optimization depends greatly on the interplay among the constituents and on the goals of the application.

With EDA, many novel strategies can be devised to deal with constraints. For example, if "sleep and wakeup strategy" is used for avoiding wasting energy, then one may have to devise an optimal schedule for nodes and/or apply a centralized strategy for both local and global constituents.

## V. RELATED WORK

Most current energy minimization approaches considered WSNs along the line of network layers: (1) the operating system, (2) the physical layer, (3) the MAC layer, (4) the network layer, (5) the application layer, and (6) the power harvesting layer. In this section we review related efforts in minimization of energy consumption at each layer.

At the Operating System (OS) level, two major efforts have been made in optimizing and managing energy consumption of the (sensor) system under its control. At the OS kernel level, one technique for minimizing the system energy consumption is processor scheduling with Dynamic voltage scaling (DVS)[10, 11]. The technique may be deployed to allocate CPU time to tasks and manipulates the CPU power states [12]. Parallel thread processing techniques can also be used to reduce energy consumption of the processor. For example, with a cluster-based infrastructure WSN, cluster heads collect data and execute the necessity computation operations in parallel. It was found that [13], "partitioning a computation creates a greater allowable latency per computation and allowing energy saving through frequency and voltage scaling".

At the Physical Layer, energy is consumed when the radio channel sends or receives data. The radio channel has three modes of operation: idle, sleep and active. Thus, the key to effective energy management is to switch the radio off when the radio channel is idle. To consume less energy, it is important to minimize the time the radio is in transmit and receive states and reduce the number of switching among different modes [14]. Furthermore, low-power listening approach may operate at the physical layer by periodically turning on the receiver to sample from incoming data. This duty-cycle approach reduces the idle listening overheads in the network [1].

Efficient MAC protocols efficiently arbitrate the use of the shared channel while aiming to reduce packet collision, idle listening, protocol overhead, and overhearing. TDMA-based protocols effectively avoid packet collisions, but their deployment in multi hop and ad hoc networks is very complex [1]. PAMAS protocol offers a technique for reducing collisions where the nodes can calculate the finish time of another node's data transfer. It saves its energy by turning itself off during the data transfer duration of other nodes. In [1], Halkes, Dam and Langendoen compare two MAC protocols (T-MAC, S-MAC) developed for wireless sensor networks. With S-MAC protocol, nodes can send queued frames during the sleeping time. Accordingly, the time between frame transmissions and idle listening is reduced. Nodes, however, are required to send SYNC messages at the start of a frame for synchronization. T-MAC adapts the duty cycle to the network traffic. It operates as S_MAC but it also uses a time-out mechanism for determining the end of the active period. The adaptive duty cycle reduces traffic fluctuation in both time and space and allow longer sleeping times.

At the network layer, several approaches may be adopted to increase the network lifetime. Topology control and related routing mechanism can be optimized for the purpose.

Determining the best topology among nodes in order to provide a connected network to route packets to the destination is a significant operation in WSNs. The challenges in selecting a suitable topology include: duty cycle control of redundant nodes, connectivity maintenance, self-configuration and redundancy identification in localized and distributed fashion [4]. Two significant methods for tackling these challenges are Geographic Fidelity (GAF) and Cluster-based Energy Conservation (CEC) protocols. GAF uses node's location information (as determined by a GPS) to configure redundant nodes and cluster them into small groups using localized and distributed algorithms. CEC has the same fundamental operation but it does not depend on location information. In [4], Xu et al. compared the two methods by simulation. They found that CEC consumes much less energy than GAF (about half) if the nodes are stationary. However, GAF is more efficient than CEC in high mobility environments. In [15], Le, Hoang and Poloah (2008) suggested a new approach for reducing protocol overhead created by CEC protocol and the energy consumption of GPS connected to sensors. In this approach, a Base Station informs the sensors about their cluster ID and cluster area by sending a sweeping beacon. If a node hears the beacon it can locate its cluster without the need for a GPS receiver.

Various kinds of topology such as tree, mesh, clustered, ad-hoc and others can be employed. In [16], Salheih et al. examine the influence of different type of mesh topologies on the power dissipated.

Since routing is a significant and costly task in WSNs as it plays a major role in determining the network lifetime and Al-Karaki and Kamal [17] discussed types of networks, topologues and protocols and their influences on the energy cost. SPIN (Sensor Protocols for Information via Negotiation) [18] is a routing technique based on node advertisements and nodes only need to know its one-hop neighbors but it is not suitable for applications which needs a reliable data delivery. LEACH (Low-Energy Adaptive Clustering Hierarchy) [19] is a clustered routing algorithm. In this method, the cluster-heads are responsible to relay data and control the cluster. Although LEACH is an effective technique for achieving prolonged network lifetime, scalability, and information security, LEACH does not guarantee optimum route. Directed Diffusion technique is a data centric, localized repair, multi-path delivery for multiple sources, sinks and queries [20]. Also, this method is able to find an optimal route.

Several technologies exist to extract energy from the environment such as solar, thermal, kinetic energy, and vibration energy and the network lifetime may increase by using power harvesting technologies. Weddell, Harris and White [21] explain advantages of energy harvesting systems as a ability of recharging after depletion and monitoring of energy consumption which may be required for network management algorithms.

## VI. CONCLUSION AND FUTURE WORK

Energy consumption is easily one of the most fundamental but crucial factor determining the success of the deployment of sensors and wireless sensor networks. This paper focuses on energy consumption as a performance measure and proposes a new Energy Driven Architecture (EDA) as a general model and approach for WSN deployment and development. This architecture deals with all common aspects of energy consumption in all types of WSNs and identifies constituents that play major roles in EDA. Designing wireless sensor networks with this architecture in mind will enable designers to balance the energy dissipation and optimize the energy consumption among all network constituents and sustain the network lifetime for the intended application.

To fulfill this goal, several important issues are being considered in our future research. One issue is to come up with a single overall formulation of the energy consumption of the entire wireless sensor network. A feasible approach, which is being explored in our next step, is to express the overall energy consumption as a linear combination of its constituent energy consumptions. Interplay among the components can be taken into account in terms of their weights as some function of the design of the WSN and the application. Other realistic but more difficult formulation expresses the energy consumption model as a non-linear function of its constituents. This approach requires more extensive exploration as we do not understand enough the metric associated with the energy of each constituent and we are unsure about the mathematical models that can handle such a non-linear relationship. Regardless of the approach taken, the aim of the application has to be taken into account as this will determines the "shape" of the overall energy consumption. For example, the requirements of the application may dictate the topology of the deployed sensor network, its routing mechanisms, or even the characteristics of the employed sensors.

Another important issue, which is being pursued in the next stage of our research, is to model comprehensively components of each of the five energy constituents of the architecture. The aim is to provide an accurate account of all functional aspects of a constituent and their salient energy-wise parameters. These parameters will allow us to evaluate the performance of WSNs, optimize their operations, and design more energy-efficient applications.

It should be noted that existing approaches for minimizing various individual components of a WSN can be adapted and/or integrated to this constituent-based architecture.